# Bayesian block-diagonal variable selection and model averaging

O. Papaspiliopoulos

ICREA and Department of Economics and Business, Universitat Pompeu Fabra

D. Rossell

Department of Economics and Business, Universitat Pompeu Fabra *

January 2, 2017

## Abstract

We propose a scalable algorithmic framework for exact Bayesian variable selection and model averaging in linear models under the assumption that the Gram matrix is block-diagonal, and as a heuristic for exploring the model space for general designs. In block-diagonal designs our approach returns the most probable model of any given size without resorting to numerical integration. The algorithm also provides a novel and efficient solution to the frequentist best subset selection problem for block-diagonal designs. Posterior probabilities for any number of models are obtained by evaluating a single one-dimensional integral that can be computed upfront, and other quantities of interest such as variable inclusion probabilities and model averaged regression estimates by carrying out an adaptive, deterministic one-dimensional numerical integration. The overall computational cost scales linearly with the number of blocks, which can be processed in parallel, and exponentially with the block size, rendering it most adequate in situations where predictors are organized in many moderately-sized blocks. For general designs, we approximate the Gram matrix by a block-diagonal using spectral clustering and propose an iterative algorithm that capitalizes on the block-diagonal algorithms to explore efficiently the model space. All methods proposed in this article are implemented in the R library mombf.

## 1 Introduction

Consider the normal linear regression model $y \sim \mathcal{N}(X\beta, \phi I)$ where $y \in \mathbb{R}^n$ is the observed outcome vector, $X$ an $n \times p$ design matrix with predictors, $\beta = (\beta_1, \ldots, \beta_p)^T \in \mathbb{R}^p$ the regression coefficients and $\phi \in \mathbb{R}^+$ the residual variance. We are interested in a sparse reconstruction of $\beta$ using an $l_0$ penalty on its entries. We follow a Bayesian formulation with spike-and-slab prior on $(\beta, \phi)$ that assigns prior probabilities on variable inclusion indicators, $\gamma = (\gamma_1, \ldots, \gamma_p)$ where $\gamma_j = \mathrm{I}(\beta_j \neq 0)$, although some of the ideas we develop apply more generally to other approaches that are based on $l_0$ penalties. For concreteness, we refer to best subset selection as the maximization of $\mathcal{N}(y; X\beta, \phi I)$ over $\beta$ under the constraint that at most $k$ elements of $\beta$ are

---

non-zero. This is typically combined with a criterion for choosing among best subsets. We refer to Bayesian best subset selection as the solutions

$$\gamma^{*m} = \arg \max_{\gamma:|\gamma|=m} p(\gamma|y,\omega), \quad m = 1, \ldots, p \qquad (1)$$

and the posterior model mode

$$\gamma^* = \arg \max_{\gamma} p(\gamma|y,\omega). \qquad (2)$$

This problem is known to be NP-hard [see for example the discussion in 4].

We provide computationally efficient algorithms for Bayesian best subset selection, inference, and model averaging under a structural relaxation of the problem, which assumes that the Gram matrix $X^T X$ is block-diagonal. The case of orthogonal design, i.e., when $X^T X$ is diagonal, has been studied extensively, and arises in factor models and wavelet regression, among others. It is well known [e.g. Proposition 3 in 4] that best subset selection becomes trivial in the orthogonal design; the regression model is also known as the Gaussian sequence model in this case, see e.g. [6]. Assuming standardised columns for simplicity, the optimal solution is $\beta_j = x_j^T y$ for all $j$ that correspond to the $k$ largest $|x_j^T y|$, $\beta_j = 0$ for all other $j$. Bayesian best subset selection in the orthogonal design has been studied before, and Section 3.3 reviews the related literature. Although important progress has been made, which this article builds upon, all previous works are unsatisfactory with respect to the treatment of $\phi$, either by fixing it to some empirical Bayes estimate or by resorting to costly Monte Carlo computations to integrate it out. Statistical properties of spike-and-slab priors have also been carefully studied, e.g. [6], however under the assumption that $\phi$ is known. The block-diagonal design is much less studied and to the best of our knowledge this is its first systematic treatment. This design is significantly harder than the orthogonal, for example best subset selection is not straightforward anymore due to the challenge of combining evidence across blocks. We provide a complete solution to Bayesian variable selection and model averaging under block-diagonal $X^T X$ that properly integrates the uncertainty about $\phi$, with computational complexity that scales linearly with the number of blocks and exponentially in the maximal block size. Central parts of the solution are a new identity for the efficient computation of the marginal posterior density of $\phi$, and an efficient algorithm for Bayesian best subset selection that does not resort to any numerical integration. For orthogonal designs the subset selection algorithm is as straightforward as the one in the frequentist formulation, while integrating out uncertainty about $\phi$. For block-diagonal designs the algorithm, which we name Coolblock, is based on a strategic use of $\phi$ used as a cooling variable to relax the optimisation problem and distribute it across blocks. It appears that a modification of Coolblock is the only available algorithm for efficient best subset selection in the block-diagonal design even in the frequentist formulation. Using the above ingredients inference on the coefficients only requires an additional one-dimensional numerical integration that can be carried out efficiently using an adaptive algorithm we develop in this article.

We showcase the potential of this machinery in general designs where $X^T X$ does not have a block structure, as an effective way to identify sets of interesting models of different sizes. The approximation of $X^T X$ by a block-diagonal matrix is linked to the so-called mincut problem, a relaxation of which is the so-called spectral clustering, which we employ to define blocks of covariates. We propose an iterative block-search algorithm that is based on model expansion and reduction steps facilitated by the use of Coolblock. We showcase the potential of this approach in approximating Bayesian best subsets when $p > n$. Our approach has conceptual links with an emerging literature in high-dimensional statistics that uses block-diagonal approximations as a "divide-and-conquer" tool; apart from works mentioned in some detail in the article we point to a recent INRIA technical report by Devijver, E. and Gallopin, M. (arxiv:1511.04033) and references therein for block-diagonal approximation in the context of graphical model selection.

## 2 Bayesian variable selection framework and previous work

We denote by $|\gamma| = \sum_{j=1}^{p} \gamma_j$ the number of variables in model $\gamma$, and define more generally for any binary vector $b$ its norm $|b|$ as the number of 1's in the vector. $X_\gamma$ is the corresponding $n \times |\gamma|$ submatrix of $X$ and $\beta_\gamma \in \mathbb{R}^{|\gamma|}$ are the regression coefficients. More generally, for any vector $\beta$ and matrix $X$, indexing by a binary vector $b = (b_1, \ldots, b_p)$ corresponds to selecting the elements in $\beta$ or columns in $X$ corresponding to $b_j = 1$ to form a new vector and matrix respectively. Finally, $x_j$ will denote the $j$'th column of $X$. Our basic assumption is that $x_i^T x_j = 0$ if variables $(i, j)$ belong to different blocks. Throughout the paper the null model for which $\gamma_j = 0$ for all $j$, will be denoted by $\emptyset$.

We let $K$ be the number of blocks and $k = 1, \ldots, K$ the block labels. The orthogonal design corresponds to $K = p$. For block $k$, $b[k]$ is defined to be a binary vector that indicates which variables are in the block, hence $b[k]_j = 1$ if and only if variable $j$ is in block $k$. We drop the block label index $k$ when not necessary and write simply $b$. Throughout $B$ denotes the maximum block size. For any two binary vectors $b, \gamma$, the product $b\gamma$ denotes element-wise multiplication. Hence, $X_{\gamma b}$ and $\beta_{\gamma b}$ denote the active variables in block $b$ and their coefficients under model $\gamma$. We typically omit $X$ as an argument in functions, since it is always conditioned upon its observed value. We consider the following Bayesian variable selection model:

$$y|\gamma, \beta, \phi \sim \mathcal{N}(X_\gamma \beta_\gamma, \phi I)$$

$$p(\beta_\gamma, \phi, \gamma \mid \tau, \rho) = \text{IG}\left(\phi; 0.5 a_\phi, 0.5 l_\phi\right) \prod_{k=1}^{K} \phi^{-\frac{|\gamma b[k]|}{2}} p(\beta_{\gamma b[k]} \phi^{-1/2} \mid \tau) \prod_{j=1}^{p} p(\gamma_j \mid \rho).$$

The first term sets an inverse gamma prior on $\phi$, a popular choice that includes improper priors as limiting cases; $a_\phi, l_\phi \in \mathbb{R}^+$ are known constants. The second term specifies a prior for the regression parameters that factorizes according to the blocks in the Gram matrix, and in which $\phi$ acts as a scale parameter. Scaling the coefficients by $\phi$ is common in this framework and can be argued in terms of invariance principles [3] and leads to more convenient computations. Common specifications include Zellner's prior

$$p(\beta_{\gamma b} \mid \phi, \tau) = \mathcal{N}\{\beta_{\gamma b}; 0, \phi, \tau(X_{\gamma b}^T X_{\gamma b})^{-1}\},$$

valid under the assumption that $X_b^T X_b$ is invertible for each block, the Gaussian shrinkage prior,

$$p(\beta_{\gamma b} \mid \phi, \tau) = \mathcal{N}(\beta_{\gamma b}; 0, \phi, \tau I)$$

and non-local priors, such as the product moment prior [11] that we consider in this article for the orthogonal design

$$p(\beta_j \mid \phi, \tau) = \frac{\beta_j^2 x_j^T x_j}{\phi \tau n} \mathcal{N}\left(\beta_j; 0, \frac{\phi \tau n}{x_j^T x_j}\right).$$

We assume that variable inclusion indicators are independent, possibly conditional on a hyperparameter $\rho$. For instance setting $P(\gamma_j = 1) = \rho$ leads to uniform $p(\gamma)$ for $\rho = 0.5$ or more generally the so-called Binomial prior for fixed $\rho \in (0, 1)$, and the Beta-Binomial prior when $\rho$ is Beta-distributed [15]. Throughout we treat the hyperparameters $\omega = (\tau, \rho)$ as fixed, however our model search algorithms remain valid as long as $p(\beta, \gamma)$ factors across blocks. In Section 6 we discuss how to adapt the framework when $\omega$ are assigned prior distributions.

Bayesian variable selection and model averaging depend critically (see Section 3.5 for details) upon the following two posterior marginal distributions: that of inclusion variables,

$$p(\gamma \mid y, \omega) = \int p(\gamma \mid y, \phi, \omega) p(\phi \mid y, \omega) \mathrm{d}\phi \,,$$

and that of the residual variance,

$$p(\phi \mid y, \omega) = \sum_\gamma p(\phi \mid y, \gamma, \omega) p(\gamma \mid y, \omega) \,.$$

Bayesian variable selection under orthogonal $X^T X$ is a classical problem that has been studied for the past two decades, whereas Bayesian best subset selection for more general block-diagonal $X^T X$ has remained largely unsolved. The main two computational bottlenecks are integrating $\phi$ and conducting model search. It is well known how to carry out Bayesian variable selection when $X^T X = I$ and $\phi$ is known. [7] gave simple closed-form expressions for posterior model probabilities $p(\gamma \mid y, \phi, \omega)$, marginal inclusion probabilities $p(\gamma_j = 1 \mid y, \phi, \omega)$, posterior means $E(\beta_j \mid y, \phi, \omega)$ and variances $\mathrm{var}(\beta_j \mid y, \phi, \omega)$ and Bayesian best subset selection conditionally on $\phi$. More recently [6] studied properties of Bayes estimators in this context. Obtaining the corresponding quantities while marginalising over $\phi$ has not had a simple exact solution to date. Previous strategies include importance sampling based on first-order Taylor approximations to the integral in $p(y \mid \gamma, \phi, \omega)$ [8], block Gibbs and other Markov chain Monte Carlo methods [7, 16], or plugging-in an estimate for $\phi$ [2, 12]. In this article we integrate $\phi$ exactly in a fast deterministic fashion and that remains applicable to block-diagonal $X^T X$. Regarding model search, Bayesian best subset selection for block-diagonal designs has received little attention. We show, to our knowledge for the first time, how to solve efficiently (1)-(2) and obtain posterior model probabilities, marginal inclusion probabilities and posterior estimates in this setting. We also illustrate how these best subset methods can be combined with covariate clustering methods to deliver efficient block-search algorithms for general $X^T X$.

# 3 Methodology

## 3.1 Fast computation of marginal distributions

The likelihood of the data given all model parameters can be factorized as follows by exploiting the block-diagonal structure of the Gram matrix:

$$p(y \mid \beta, \gamma, \phi) = (2\pi\phi)^{n/2} \exp\left(-\frac{1}{2\phi} y^T y\right)$$

$$\times \exp\left\{-\frac{1}{2\phi} \sum_{k=1}^{K} \left(\beta_{\gamma b[k]}^T X_{\gamma b[k]}^T X_{\gamma b[k]} \beta_{\gamma b[k]} - 2y^T X_{\gamma b[k]} \beta_{\gamma b[k]}\right)\right\}.$$

Using the corresponding factorisation of the prior on the regression coefficients we obtain

$$p(y \mid \gamma, \phi, \tau) = \exp\left(-\frac{1}{2\phi} y^T y\right) (2\pi\phi)^{-\frac{n}{2}} \prod_{k=1}^{K} f(y\phi^{-1/2}, \gamma b[k], \tau). \tag{3}$$

where for convenience we use the change of variables $\tilde{\beta}_\gamma = \beta_\gamma \phi^{-1/2}$ to define

$$f(z, \gamma, \tau) = \int e^{z^T X_\gamma \tilde{\beta}_\gamma} e^{-\frac{1}{2} \tilde{\beta}_\gamma^T X_\gamma^T X_\gamma \tilde{\beta}_\gamma} p(\tilde{\beta}_\gamma \mid \tau) \mathrm{d}\tilde{\beta}_\gamma \text{ for } |\gamma| > 0 \quad f(z, \emptyset, \tau) = 1. \tag{4}$$

The factorisation (3) generalizes a well-known result for the orthogonal design [7], and leads to the factorisation of posterior model probabilities given $\phi$

$$p(\gamma \mid y, \phi, \omega) \propto \prod_{k=1}^{K} f(y\phi^{-1/2}, \gamma b[k], \tau) p(\gamma b[k] \mid \rho) \propto \prod_k p(\gamma_{b[k]} \mid y, \phi, \omega). \tag{5}$$

The last proportionality acknowledges that each term in the product can be normalized to yield a joint probability density for the variables within each block. The normalisation simply involves a summation over $2^{|b[k]|}$ terms, for each block $k$. We now note that by Bayes theorem we obtain the alternative expression

$$p(\gamma \mid y, \phi, \omega) = \frac{p(y, \gamma \mid \phi, \omega)}{p(y \mid \phi, \omega)},$$

which after re-arrangement and combining with (3) and (5) gives

$$p(y \mid \phi, \omega) = \exp\left(-\frac{1}{2\phi} y^T y\right) (2\pi\phi)^{-\frac{n}{2}} \prod_{k=1}^{K} \sum_{\gamma b[k]} f(y\phi^{-1/2}, \gamma b[k], \tau) p(\gamma b[k] \mid \rho), \tag{6}$$

where the sum in each product is effectively over $2^{|\gamma b[k]|}$ terms, hence at most $2^B$ terms. Exploiting again the same ingredients and the inverse gamma prior on $\phi$, we obtain $p(y \mid \gamma, \omega) =$

$$\frac{\left(\frac{l_\phi}{2}\right)^{\frac{a_\phi}{2}}}{(2\pi)^{\frac{n}{2}} \Gamma\left(\frac{a_\phi}{2}\right)} \times \int \left(\frac{1}{\phi}\right)^{\frac{a_\phi + n}{2} + 1} \exp\left\{-\frac{1}{2\phi}\left(l_\phi + y^T y\right)\right\} \prod_{k=1}^{K} f(y\phi^{-1/2}, \gamma b[k], \tau) \mathrm{d}\phi. \tag{7}$$

Section 3.2 provides simple formulae for $f$ and the two marginal likelihoods (6)-(7) for the main examples we consider in this paper. Expression (6) has the key implication that $p(y \mid \phi, \omega)$ can be evaluated pointwise by computing a simple product, avoiding a cumbersome summation with $2^p$ terms over the model space. Although trivial, to our knowledge this factorization has not been exploited by earlier computational schemes. Expression (7) is useful but less critical as for many prior choices this marginal likelihood often has closed-form or can be quickly evaluated.

## 3.2 Some examples

To gain intuition we start with specific examples of the function $f$ defined in (4) for some priors commonly used in the literature. We also provide expressions for $p(y \mid \gamma, \omega)/p(y \mid \emptyset, \omega)$, useful to compare which amongst any two models has higher posterior probability, which combined with the search algorithms in Section 3.3-3.4 allows determining the posterior mode.

In the case of Zellner's prior a direct calculation gives the well-known result

$$f(z, \gamma b[k], \tau) = (\tau + 1)^{-\frac{|\gamma b[k]|}{2}} \exp\left\{\frac{1}{2}\frac{\tau}{\tau+1}u(z, \gamma b[k])\right\}, \tag{8}$$

where $u(z, \gamma) = z^T X_\gamma (X_\gamma^T X_\gamma)^{-1} X_\gamma^T z$ with the convention that $u(z, \emptyset) = 0$. The statistic $u(y, \gamma)$ evaluated at $z = y$ is proportional to the squared Mahalanobis distance between the least-squares estimator of $\beta_\gamma$ (equivalently, its posterior mean under Zellner's prior) and 0. Note also that a direct implication of the block-diagonal assumption is that

$$\sum_k u(z, \gamma b[k]) = u(z, \gamma)$$

a property we exploit below to provide compact expressions. Then (7) expressed as the ratio of integrated likelihoods relative to the null model becomes

$$\frac{p(y \mid \gamma, \omega)}{p(y \mid \emptyset, \omega)} = \left(\frac{l_\phi + y^T y}{l_\phi + y^T y - \frac{\tau}{\tau+1}u(y, \gamma)}\right)^{\frac{a_\phi + n}{2}} \frac{1}{(1+\tau)^{|\gamma|/2}}. \tag{9}$$

For the product moment prior in the orthogonal design case $K = p$ (thus $|\gamma b[k]| = 1$) we obtain

$$f(z, \gamma b[k], \tau) = (\tau n + 1)^{-\frac{3}{2}} \exp\left\{\frac{1}{2}\frac{\tau n}{\tau n + 1}u(z, \gamma b[k])\right\}\left\{\frac{\tau n}{(\tau n + 1)}u(z, \gamma b[k]) + 1\right\}$$

which resorting to the multi-binomial theorem leads to

$$\frac{p(y \mid \gamma, \omega)}{p(y \mid \emptyset, \omega)} = \left(\frac{l_\phi + y^T y}{l_\phi + y^T y - \frac{\tau n}{\tau n + 1}u(y, \gamma)}\right)^{\frac{a_\phi + n}{2}} \frac{1}{(1+\tau n)^{|\gamma|/2}} \times$$

$$\sum_{a_1=0}^{1} \cdots \sum_{a_{|\gamma|}=0}^{1} \frac{\Gamma\left(\frac{a_\phi + n}{2} + |\gamma| - |a|\right)}{\left\{l_\phi + y^T y - \frac{\tau n}{\tau n + 1}u(y, \gamma)\right\}^{|\gamma|-|a|} \Gamma\left(\frac{a_\phi + n}{2}\right)}. \tag{10}$$

The summation is over $2^{|\gamma|}$ terms, thus for large $|\gamma|$ it is more efficient to evaluate the corresponding $p(\gamma \mid y, \omega)$ not by using the above expression, but instead via $p(\gamma \mid y, \omega) = \int p(\gamma \mid y, \phi, \omega) p(\phi \mid y, \omega) \mathrm{d}\phi$ as described in Section 3.5. For the Gaussian shrinkage prior defined earlier we get

$$f(z, \gamma b[k], \tau) = \tau^{-\frac{|\gamma b[k]|}{2}} |X_{\gamma b[k]}^T X_{\gamma b[k]} + \tau^{-1} I|^{-1/2} \exp\left\{\frac{1}{2} s(z, \gamma b[k])\right\}$$

where $s(z, \gamma b[k]) = z^T X_{\gamma b[k]} (X_{\gamma b[k]}^T X_{\gamma b[k]} + \tau^{-1} I)^{-1} X_{\gamma b[k]}^T z$. Direct algebra also gives the well-known result

$$\frac{p(y \mid \gamma, \omega)}{p(y \mid \emptyset, \omega)} = \left(\frac{l_\phi + y^T y}{l_\phi + y^T y - s(y, \gamma)}\right)^{\frac{a_\phi + n}{2}} \tau^{-|\gamma|/2} |X_\gamma^T X_\gamma + \tau^{-1} I|^{-1/2}$$

where as before $s(z, \gamma) = \sum_k s(z, \gamma b[k])$ and we have exploited a similar property for the log-determinant.

### 3.3 Bayesian best subset selection for orthogonal designs

Section 3.1 obtains expressions for $p(y \mid \gamma, \omega)$, hence up to normalising constant for $p(\gamma \mid y, \omega)$. However, identifying $\gamma^{*m}$ for all $m = 1, \ldots, p$, as defined in (1), still in principle requires full enumeration of $2^p$ models. The marginal with respect to $\phi$ posterior model probabilities, $p(y \mid \gamma, \omega)$, do not admit a simple factorisation. On the other hand, under an assumption on the dependence of $f(z, \gamma, \tau)$ on the $u(z, \gamma)$-scores, a simple algorithm for identifying $\gamma^{*m}$ is available, already pointed out by [7] for Zellner's prior and effectively coinciding with frequentist best subset selection as described in Section 2. Our result applies to a more general class of priors satisfying the assumption below.

**Assumption 3.1.** The function $f$ in (4) can be expressed as $f(z, \gamma, \tau) = h\{u(z, \gamma), |\gamma|, \tau\}$ where $h$ is non-decreasing in its first argument, and $u(z, \gamma)$ as in (8). □

Zellner's prior (and scale mixtures thereof) clearly satisfies Assumption 3.1. For diagonal $X^T X$, and provided that covariates have been standardized to unit variance, i.e. $x_j^T x_j = n$, Anderson's lemma ([1], Lemma 10.2 and Remark 10.1, pp. 157) implies that Assumption 3.1 is met whenever $p(\beta_j \mid \tau)$ is symmetric around 0 and $p(\beta_j \mid \tau)/\mathcal{N}(\beta_j; 0, 1)$ is increasing in $|\beta_j|$. Thus for orthogonal designs a sufficient condition is that $p(\beta_j \mid \tau)$ is centred at 0 and has normal or thicker-than-normal tails, as is the case for the vast majority of spike-and-slab priors in the literature, including Normal, T and all non-local priors proposed to date. More generally, the assumption asks that $f$, which is a marginal likelihood quantity, depends on the data only via the goodness of fit statistic $u(y, \gamma)$ in an non-decreasing way, and on $\gamma$ only through its size $|\gamma|$ (typically it will be non-increasing in $|\gamma|$). That is, the farther $E(\beta_\gamma \mid y, \gamma)$ is from 0 the higher are the odds in favour of $\gamma$ - this seems reasonable provided of course that the Mahalanobis distance to 0 (equivalently, the sum of squared residuals) is an appropriate metric for the deviation from the null. As an example where this does not happen, under the Gaussian shrinkage prior for non-orthogonal designs both the goodness of fit and the volume of $X_\gamma^T X_\gamma$ are used to compute $f$.

**Proposition 3.2.** Let $K = p$, $r_j = |x_j^T y|/(x_j^T x_j)^{\frac{1}{2}}$ and $J$ be the set of indices of the $m$ largest scores $r_j$. Then $\gamma_i^{*m} = 1$ for all $j \in J$ and $\gamma_i^{*m} = 0$ for $j \notin J$.

*Proof.* In the orthogonal design $u(z, \gamma) = \sum_{i:\gamma_i \neq 0} r_i^2$, hence $u(z, \gamma) \leq \sum_{j \in J} r_j^2$, hence $u(y\phi^{-1/2}, \gamma) \leq u(y\phi^{-1/2}, \gamma^{*m})$ for $\gamma^{*m}$ as defined above, and for all $\gamma$ with $|\gamma| = m$. Assumption 3.1 then implies that $f(y\phi^{-1/2}, \gamma, \tau) \leq f(y\phi^{-1/2}, \gamma^{*m}, \tau)$ for all $\gamma$ with $|\gamma| = m$, and for $c$ the normalising constant,

$$p(\gamma \mid y, \phi, \omega) = cp(1 \mid \rho)^m p(0 \mid \rho)^{p-m} f(y\phi^{-1/2}, \gamma, \tau)$$
$$\leq cp(1 \mid \rho)^m p(0 \mid \rho)^{p-m} f(y\phi^{-1/2}, \gamma^{*m}, \tau) = p(\gamma^{*m} \mid y, \phi, \omega).$$

The result follows by integrating both sides with respect to $p(\phi|y, \omega)$. $\square$

This result becomes particularly useful for Bayesian best subset selection when combined with the results of Section 3.1 which provide a fast computation of $p(\gamma \mid y, \omega)$ up to normalising constant for each model $\gamma$. We can efficiently identify $\gamma^{*m}$ for every $m$ by means of Proposition 3.2 and then find $\gamma^*$ by comparing the unnormalized posterior probabilities computed efficiently as described in Section 3.1. Of course, to assign probabilities to each of these models we need to compute the normalising constant $\sum_\gamma p(y \mid \gamma, \omega) p(\gamma \mid \rho)$ and this is addressed in Section 3.5.

## 3.4 Bayesian best subset selection for block-diagonal designs with Zellner's prior

Under orthogonal design the solution of (1) is straightforward under very mild assumptions on $p(\beta \mid \gamma)$. This is not the case with block-diagonal designs when $K < p$. Throughout this section we focus on Zellner's prior and discuss extensions in Section 6. The main idea is to connect best subset selection under $p(\gamma \mid y, \omega)$ with that under $p(\gamma \mid y, \phi, \omega)$. Selection under the latter, when $\phi$ is known, is easy and can be distributed across blocks.

$$\gamma^{*\phi} = \arg\max_\gamma p(\gamma \mid y, \phi, \omega)$$

for the model with the highest posterior probability conditionally on a given $\phi$. The methodology we propose is based on the following observation that the best model for a given $\phi$ is also the best model amongst all models of its same size.

**Proposition 3.3.** *If $p(\beta_\gamma \mid \gamma, \omega)$ is Zellner's prior and for some given $\phi$, $|\gamma^{*\phi}| = m$, then $\gamma^{*\phi} = \gamma^{*m}$.*

*Proof.* From (5) and (8) we have that

$$p(\gamma \mid y, \phi, \omega) \propto p(1 \mid \rho)^{|\gamma|} p(0 \mid \rho)^{p-|\gamma|} (\tau+1)^{-|\gamma|/2} \exp\left\{\frac{1}{2}\frac{\tau}{\tau+1}\frac{1}{\phi} u(y, \gamma)\right\}$$

hence, $\gamma^{*\phi}$ has the highest $u(y, \gamma)$-score among all models $\gamma$ of size $m$, which according to (9) implies that $\gamma^{*\phi}$ has the highest marginal posterior probability among all models of size $m$. $\square$

This proposition opens the possibility of exploiting optimisation conditionally on $\phi$ in order to identify highest posterior probability models marginally with respect to $\phi$ of different sizes, which would otherwise require solving a hard optimisation problem. In particular,

$$\gamma^{*\phi} = \arg\max_\gamma \prod_{k=1}^K f(y\phi^{-1/2}, \gamma b[k], \tau) p(\gamma b[k] \mid \rho) \tag{11}$$

can be found by maximising each term in the product separately, which can be distributed across $K$ decoupled optimisations. Our key finding, detailed below, is a closed-form expression for a series of values $\{\phi_t\}$ for $t = 0, \ldots, T$, where $T = \mathcal{O}(p)$, such that the corresponding posterior modes range from the null to the full model.

To achieve this, we first work within each block $k$. By fully enumerating the $2^{|b[k]|}$ variable configurations we do within-block best subset selection. We write (in accordance with the notation used so far)

$$\gamma b[k]^{*m} = \arg \max_{\gamma : |\gamma b[k]| = m} u(y, \gamma b[k]), m = 1, \ldots, |b[k]|. \quad (12)$$

the best $m$-variable configuration within block $k$. We next compare these configurations between different sizes $l > j$, noting that $p(\gamma b[k]^{*l} \mid y, \phi, \omega) \geq p(\gamma b[k]^{*j} \mid y, \phi, \omega)$ if and only if

$$\frac{\tau}{\tau+1}\frac{1}{\phi}\{u(y, \gamma b[k]^{*l}) - u(y, \gamma b[k]^{*j})\} \geq (l-j)\log(1+\tau) + 2\log\left\{\frac{p(\gamma b[k]^{*j} \mid \rho)}{p(\gamma b[k]^{*l} \mid \rho)}\right\}. \quad (13)$$

Provided that prior model probabilities are non-increasing in model size (as is normally the case, see examples in Section 2) both sides of the above inequality are positive. For the prior $\prod_{j=1}^{p} p(\gamma_j \mid \rho)$ we have assumed in the basic formulation, the right hand side in (13) becomes

$$(l-j)\{\log(1+\tau) + 2\log[p(0 \mid \rho)/p(1 \mid \rho)]\}.$$

The root of (13) in terms of $\phi$ gives the level for the residual variance, which we will call $r_{klj}$, such that when $\phi \leq r_{klj}$ the larger model is preferable in terms of its conditional posterior probability, whereas for $\phi > r_{klj}$ the smaller model is prefered. We also define $r_{kjl} = r_{klj}$ when $j < l$. We now state a basic but important characterisation of within-block optimality.

**Proposition 3.4.** *The variable configuration $\gamma b[k]^{*l}$ as defined in* (12) *has the highest conditional posterior probability $p(\gamma b[k] \mid y, \phi, \omega)$ among all variable configurations $\gamma b[k]$ from block $k$, provided* $\max_{j>l} r_{klj} \leq \phi < \min_{j<l} r_{klj}$.

The proof follows directly from (13). The interval defined in the proposition might be empty, in which $\gamma b[k]^{*l}$ is not preferrable over all other models for any value of $\phi$. The implication of this result is the identification of a sequence of critical $\phi$ values and a sequence of variable configurations of increasing size for each block, as described in Algorithm 1. For each block $k = 1, \ldots, K$ the algorithm returns a subset of the best variable configurations of sizes $j = 1, \ldots, |b[k]|$, whose defining characteristic is that for a certain range of $\phi$ values they are preferable to all other within-blocks configuration. More precisely, let $\phi_{kt}$ be the $t$'th $\phi$ value in order of addition to the $\Phi_k$ list, where by construction the $\phi_{kt}$'s are in decreasing order with respect to the index $t$. Then, the $t$'th corresponding variable configuration has the highest conditional posterior probability among all configurations when $\phi \in (\phi_{k(t+1)}, \phi_{kt})$. Hence the variable configurations not included in the list output by Algorithm 1 are always dominated by some other combination of variables for any given $\phi$. We feed this output into Algorithm 2, which we nicknamed "Coolblock", which then combines variable configurations across blocks by "cooling" $\phi$ (from larger to smaller values) to identify the highest marginal probability models of different sizes.

**Algorithm 1.** Enumblock (applied to block $k$)

    **Data:** $\{r_{klj}; l, j = 1, \ldots, |b[k]|\}, \{\gamma b[k]^{*j}, j = 1, \ldots, |b[k]|\}$
    **Result:** A list of $\phi$-values $\Phi_k$ and a list of models $\Gamma_k$
    Initialisation: set $\Phi_k = []$ and $\Gamma_k = []$ to be empty lists;
    $J = [0]$ is a list containing the element 0;
    $j = 1$;
    **while** $(j < |b[k]|)$ **do**
        **if** $\min_{l \in J} r_{klj} > \max_{l > j} r_{klj}$ **then**
            append $\min_{l \in J} r_{klj}$ to $\Phi_k$ and $\gamma b[k]^{*j}$ to $\Gamma_k$;
            append $j$ to $J$;
        **end**
        $j = j + 1$;
    **end**
    append $\min_{l < j|} r_{klj}$ to $\Phi_k$ and $\gamma b[k]^{*j}$ to $\Gamma_k$ ;

**Algorithm 2.** Coolblock

    **Data:** Variable configurations $\Gamma_k$ and residual variances $\Phi_k$ output by Enumblock
    **Result:** A list $M$ of models of different sizes containing the highest marginal posterior probability
                among all models of their size
    Initialisation: $M = [\emptyset], \gamma = (0, \ldots, 0)$;
    Concatenate $\Phi = \bigcup_{k=1}^{K} \Phi_k$, $\Gamma = \bigcup_{k=1}^{K} \Gamma_k$ and let $\kappa = (1, \ldots, 1, 2, \ldots, 2, \ldots, K, \ldots, K)$ be the block
    indexes for the elements in $\Phi, \Gamma$.
    Order elements of $\Phi$ in decreasing order. For any $\phi \in \Phi$ let $\kappa(\phi)$ and $\Gamma(\phi)$ be the corresponding
    elements in $\kappa, \Gamma$;
    **for** $\phi \in \Phi$ **do**
        $\gamma b[\kappa(\phi)] = \Gamma(\phi)$
    **end**
    Append $\gamma$ to $M$

By construction, for each $\phi \in \Phi$, the model $\gamma$ that maximizes $p(\gamma \mid y, \phi, \omega)$ is found by optimising each term in the product (11) separately, i.e. choosing from each block the variable configuration that it is optimal for this $\phi$. Given that elements in $\phi \in \Phi$ correspond to levels of residual variance at which the optimal choice changes for some block $\kappa(\phi)$, "Coolblock" simply replaces the configuration in that block with its new optimal configuration (denoted $\Gamma(\phi)$ in the algorithm) and keeps the rest of the blocks unchanged. By virtue of Proposition 3.3 each model in $\gamma \in M$ has the largest $p(\gamma \mid y, \omega)$ among all models of its size. By construction the models in $M$ are ordered in increasing size.

We remark that "Coolblock" outputs a series of models $M$, each of which has largest $p(\gamma \mid y, \omega)$ amongst all models of its size, however not all model sizes need be contained in $M$. This arises from the fact that for strongly correlated covariates the optimal choice as $\phi$ decreases may be to increase model size by more than 1 variable, and is formally connected to the $\min_{l \in J} r_{klj} > \max_{l > j} r_{klj}$ condition in Algorithm 1. We do not view this issue as problematic. For any model $\gamma \notin M$, by construction for all $\phi$ there exists some $\gamma' \in M$ such that $p(\gamma \mid y, \phi, \omega) < p(\gamma' \mid y, \phi, \omega)$. Hence any $\gamma \notin M$ is unlikely to be the mode marginally across $\phi$, as this would require $\gamma$ to be better than each $\gamma' \in M$ for a certain subset of $\phi$ values, and that the intersection of all these subsets is the empty set. However, for the sake of completeness we outline how to formally evaluate the possibility that a model outside $M$ is the mode. A quick test for such a situation is facilitated by using an upper bound. Let $\gamma \notin M$ and $\gamma' \in M$ the smallest model in $M$ with larger size than $\gamma$. Then $u(y, \gamma) \leq u(y, \gamma')$, since the residual sum of squares for the best size $|\gamma'|$ model is smaller than for the best size $|\gamma|$ model, giving that

$$\frac{p(\gamma \mid y, \omega)}{p(\emptyset \mid y, \omega)} \leq \frac{p(\gamma \mid \rho)}{p(\emptyset \mid \rho)} \left\{ \frac{l_\phi + y^T y}{l_\phi + y^T y - \frac{\tau}{\tau+1} u(y, \gamma')} \right\}^{\frac{a_\phi + n}{2}} \frac{1}{(1+\tau)^{|\gamma|/2}}. \tag{14}$$

Thus if the right-hand-side upper bound is smaller than the corresponding ratios computed for models in $M$ then $\gamma$ cannot be the mode. As illustrated in our examples this simple test typically rules out most model sizes as candidates for the posterior mode, occasionally leaving a few model sizes $|\gamma|$, where $|\gamma|$ is usually small. For these remaining sizes we resort to exhaustive model enumeration, i.e. consider all possible ways to allocate the $|\gamma|$ variables to the $K$ blocks. The number of such combinations is greatly constrained by the capacity of each block, but to gain intuition a simple upper bound is obtained by the unconstrained case (all blocks have $> |\gamma|$ variables). The "stars and bars" method in Section II.5 of [9] yields that the unconstrained number of allocations is $|\gamma| + K - 1$ choose $K - 1$ which is orders of magnitude smaller than the $p$ choose $|\gamma|$ in general. For each permissible allocation schedule, say $(s_1, \ldots, s_K)$ that includes $s_k \geq 0$ variables from block $k$, by construction the optimal model is $\sum_k \gamma^*_{s_k} b[k]$. Thus, given the work already done, no further search is needed. We only need to compute $p(\sum_k \gamma b[k]^{*s_k} \mid y, \omega)$ up to proportionality for each allocation schedule. The supremum over those is the best size $|\gamma|$ model.

### 3.5 Numerical integration with respect to $\phi$

Let $g(\phi)$ be a quantity of interest that is available for given $\phi$, we now discuss how to evaluate

$$E(g(\phi) \mid y, \omega) = \int g(\phi) p(\phi \mid y, \omega) d\phi. \tag{15}$$

Specific instances worth pointing out are posterior model probabilities

$$g(\phi) = p(\gamma \mid y, \phi, \omega) = \prod_{k=1}^{K} \frac{f(y\phi^{-1/2}, \gamma b[k], \tau) p(\gamma b[k] \mid \rho)}{\sum_{\gamma' b[k]} f(y\phi^{-1/2}, \gamma' b[k], \tau) p(\gamma' b[k] \mid \rho)},$$

marginal inclusion probabilities, say for variable $j$ in block $k$,

$$g(\phi) = \frac{\sum_{\gamma b[k]:\gamma_j=1} f(y\phi^{-1/2}, \gamma b[k], \tau) p(\gamma b[k] \mid \rho)}{\sum_{\gamma b[k]} f(y\phi^{-1/2}, \gamma b[k], \tau) p(\gamma b[k] \mid \rho)}$$

or posterior means of the regression parameters

$$g(\phi) = E(\beta_{b[k]} \mid y, \phi, \omega) = \sum_{\gamma b[k]} E(\beta_{\gamma b[k]} \mid y, \gamma, \phi, \omega) p(\gamma b[k] \mid y, \phi, \omega).$$

Our strategy to compute Expression (15) is to evaluate its integrand on an adaptive grid $\phi_t$ for $t = 0, \ldots, T$ and then using any convenient numerical integration, e.g., in our examples composite Simpson's rule and adaptive quadrature (R function integrate). The integrand requires $p(\phi \mid y, \omega)$ but from (6) only its un-normalized version $p(y \mid \phi, \omega) p(\phi)$ is available. Hence we estimate upfront the normalising constant $p(y \mid \omega)$ using numerical integration on $p(y \mid \phi, \omega) p(\phi)$ and then use the resultant $p(\phi \mid y, \omega)$ for any subsequent calculations.

We propose a strategy for choosing the grid to cope with the fact that since $p(\phi \mid y, \omega) = \sum_\gamma p(\phi \mid y, \gamma, \omega) p(\gamma \mid y, \omega)$ is a mixture it can be multimodal, although our experience in practice suggests that this does not often happen. We take advantage of the Bayesian best subset selection algorithms to set as an initial grid posterior modes $\arg\max_\phi p(\phi \mid y, \gamma^{*m}, \omega)$, for $m = 0, 1, \ldots, p$. From (7) these modes are given (or approximated) by an inverse gamma and are strictly bounded between the modes under the null and full models. Hence by setting $\phi_0$ to the 99.9% and $\phi_T$ to the 0.1% percentiles of these two models the initial grid contains the whole range of values with high $p(\phi \mid y, \omega)$. One could alternatively set this initial grid to the $\phi \in \Phi$ values returned by Algorithm 2, at no extra cost. The initial grid is then refined by sub-dividing the interval $[\phi_t, \phi_{t+1}]$ whenever $|p(\phi_t \mid y, \omega) - p(\phi_{t+1} \mid y, \omega)|$ is large, e.g., in our examples we kept subdividing until these differences were $< 0.01 \times \max_t p(\phi_t \mid y, \omega)$. In a similar fashion for computational convenience one may drop values from the initial grid when $|p(\phi_t \mid y, \omega) - p(\phi_{t+1} \mid y, \omega)|$ is negligible, which in our experience typically gives a final grid which only has a few hundred points. Altogether, from (6) each evaluation of the integrand requires $K\mathcal{O}(2^B)$ operations - to compute $p(\phi \mid y, \omega)$ and this will typically be an upper bound on the computation of $g(\phi)$ itself - thus the cost to evaluate (15) is $\mathcal{O}(TK2^B)$, where $T$ is usually in the hundreds. These operations are embarrassingly parallel.

## 4 Subset selection for general designs using block-diagonal relaxations

### 4.1 A novel block search algorithm and some related work

The Coolblock algorithm remains useful when $X^T X$ is not truly block-diagonal. As proof-of-principle we outline a novel block search algorithm that scales to $p > n$. It defines a forward-backward heuristic for optimizing $p(\gamma \mid y)$ under arbitrary designs and prior distributions. The idea is to define blocks by clustering correlated variables, use Coolblock to define a set of interesting models $M$, and do Bayesian best subset selection among the models in $M$.

"Divide-and-conquer" strategies that exploit block-diagonal approximations have been explored before, typically under the assumption of a known ordering of the covariates under which the population covariance is near-block diagonal. Then, the re-construction of the blocks is done either by tapering, see e.g. [5], or linear filtering, e.g. as in [13]. However, in generic contexts such ordering is unknown. A work with some connections to ours is in Chapter 7 of a 2013 Berkeley PhD thesis by M. Cai., who proposes creating overlapping groups of highly-correlated covariates and then a screen and clean approach, which at the screen step considers adding a whole group to an existing set of active variables.

We take a different approach here. Our aim is to identify Bayesian best subsets for general designs and priors, not just Zellner's as in Section 3. We use spectral clustering to define blocks of correlated covariates and hence approximate the Gram matrix by a block-diagonal. We describe this construction in Section 4.2. Once the blocks have been defined, we use Coolblock to identify a set $M$ of interesting subsets of different sizes. If $X^T X$ is not block-diagonal $M$ may not contain all best subsets anymore, but to the extent that $X^T X$ entries outside the inferred blocks are smaller than within-block entries, $M$ will contain high probability models. The model identified by this heuristic model search algorithm will be denoted by $\hat{\gamma}^*$, to separate it from the exact posterior mode $\gamma^*$. Taking $\hat{\gamma}^* = \arg\max_{\gamma \in M} p(\gamma \mid y, \omega)$ may lead to a mild-quality local mode. Instead, we propose the iterative scheme in Algorithm 3 that refines the model search. Step 2 drops variables from $\hat{\gamma}^*$ by defining a new sequence of models based on clustering the columns of $X_{\hat{\gamma}^*}$. Step 1 adds variables to the updated $\hat{\gamma}^*$ based on partial residuals after regressing $y$ on $X_{\hat{\gamma}^*}$. The process is repeated until the posterior probability of $\hat{\gamma}^*$ does not improve between two successive iterations. We remark that the algorithm is valid for any $p(\beta \mid \gamma)$ and $p(\gamma)$, it simply uses spectral clustering and Coolblock to identify sets of potentially interesting models, once these are available one may evaluate $p(y, \gamma \mid \omega)$ exactly for any desired priors. In other words, the algorithm does a heuristic optimisation on a well-defined objective function over the model space. Additionally, if $X^T X$ is really block-diagonal and Zellner's prior is used for the coefficient, Algorithm 3 returns the correct Bayesian best subsets.

### 4.2 Block-diagonal approximation by spectral clustering

We use graph-theory tools to decompose the set of variables into blocks. We first standardize each covariate by subtracting sample mean and dividing by sample variance. Expressions below assume that each covariate $x_j$ is thus transformed, and denote the sample covariance matrix of the covariates as $S = X^T X / n$. Then, the block diagonal matrix closest to $S$ in squared Frobenius norm, is found by solving the mincut problem associated with a weighted graph for which each node is a variable and the the weight between nodes $i$ and $j$ is $w_{ij} = (x_i^T x_j)^2$. The matrix of weights $W = S \odot S$ is a Hadamard product, which is positive semi-definite by means of the Schur's product theorem. The mincut problem is to split the nodes in a manner such that the sum of the weights across blocks is minimised. Typically we are interested in a variation of this problem that penalizes small block sizes (such as the so-called ratio cut). These problems are known to be an NP-hard. Spectral clustering solves a problem relaxation based on the $k$ smallest eigenvalues of the graph Laplacian, where $k$ is the number of desired blocks; see [17] for a review. There are several formulations of spectral clustering, here we work with

**Algorithm 3.** Block search

    **Data:** $X$, $y$, maximum block size $B$ and iterations $I$
    **Result:** A list of models $M$, $p(y, \gamma \mid \omega)$ and $\hat{\gamma}_k^* = \arg\max_{\gamma \in M : |\gamma|=k} p(\gamma \mid y, \omega)$.
    Initialisation: set $e = y$, $M = [\emptyset]$, $\hat{\gamma}^* = \emptyset$, $\hat{f}^* = p(\hat{\gamma}^*, y|\omega)$, $i = 0$, increase=true;
    **while** $(i < I)$ *and (*`increase`*)* **do**
        Step 1. Add variables to $\hat{\gamma}^*$
            Define blocks of columns in $X_{1-\hat{\gamma}^*}$ via spectral clustering;
            Run Coolblock on $(e, X_{1-\hat{\gamma}^*})$ to obtain a set of models $N$ of maximum size $\min\{n, p\}$
            Add active variables in $\hat{\gamma}^*$ to models in $N$, set $M = M \cup N$;
        $\hat{\gamma}^* = \arg\max_M p(\gamma, y \mid \omega)$
        Step 2. Drop variables from $\hat{\gamma}^*$
            Define blocks of columns in $X_{\hat{\gamma}^*}$ via spectral clustering;
            Run Coolblock on $(y, X_{\hat{\gamma}^*})$ to obtain a list of models $N$;
        $M = M \cup N$, $\hat{\gamma}^* = \arg\max_M p(\gamma, y \mid \omega)$, $f = \max_M p(\gamma, y \mid \omega)$;
        If $f = \hat{f}^*$ then increase = false;
        $\hat{f}^* = f$;
        Set $e = y - X_{\hat{\gamma}^*} \hat{\beta}_{\hat{\gamma}^*}$, where $\hat{\beta}_{\hat{\gamma}^*} = (X_{\hat{\gamma}^*}^T X_{\hat{\gamma}^*})^{-1} X_{\hat{\gamma}^*} y$
        $i = i + 1$;
    **end**

the following. We obtain the $k$ largest eigenvalues and corresponding eigenvectors of the normalized affinity matrix, $A = D^{-\frac{1}{2}} W D^{-\frac{1}{2}}$ where $D$ is a diagonal matrix containing the row sums of $W$. By construction 1 is the largest eigenvalue and its eigenvector has all entries proportional to 1. Also, if $S$ is really block-diagonal, the eigenvalue 1 has multiplicity $k$ and the associated eigenvectors indicate the block-membership structure. When $S$ does not have a block structure, the separation of nodes into blocks is done by using a $k$-means clustering algorithm on the $k$ eigenvectors of $A$ that correspond to the $k$ smallest eigenvalues. In our implementation we work with the eigenvectors scaled by the corresponding eigenvalue. To prevent a poor k-means initialization and also to provide a fully deterministic algorithm, we set the initial clusters using the $k$ quantiles of the first scaled eigenvector. To ensure that no block has size above a user-specified $B$ (in our examples $B = 10$) we set $k = p/B$ and subdivide any cluster of size larger than $B$ by iteratively applying 2-means, until all blocks have size no larger than $B$. Section 5.5 illustrates the approach.

# 5 Examples

## 5.1 Computer code and reproducibility

We illustrate our methodology in several examples, including cases where $X^T X$ is not block-diagonal. The R code is provided in a Supplement to the article, together with additional figures, and the main routines are available in the R package mombf version 1.8.3.

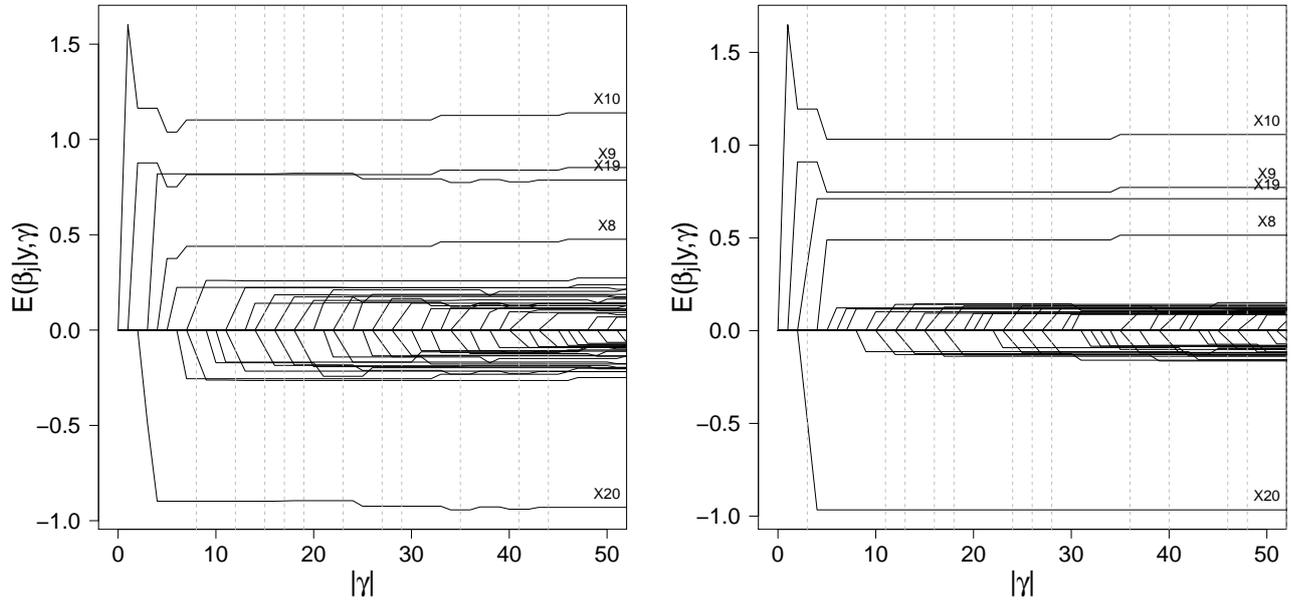

Figure 1: First 50 models identified by the "Coolblock" algorithm and corresponding $E(\beta \mid y, \gamma, \omega)$ when $p = 100, n = 150$ and $p = 500, n = 510$. Dashed lines indicate model sizes that were not visited.

## 5.2 Orthogonal design

A Supplement to this article contains experiments, and details on their implementation, with the orthogonal design when $p = 500$ and $n = 510$. As an illustration the total time on a Mac laptop (2 GHz Intel core i7 processor, 8 GB 1600 MHz DDR3, OS X 10.11.4) to find the sequence of 500 models, their posterior probabilities, and the marginal inclusion probabilities and $E(\beta_j \mid y)$ for all 500 variables was 0.3 seconds for Zellner's and 5.5 seconds for the product moment priors. We emphasize that this is a pure R language implementation, a lower-level language such as C typically cuts running times by a factor of 10 or more, and we did not use any parallel processing. It should also be noted that in practice one is often interested only in posterior probabilities and parameter estimates for the first top models or variables, requiring only a fraction of our computations.

## 5.3 Block-diagonal design

We present a block-diagonal example with $p = 100, n = 150$ and subsequently under $p = 500, n = 510$. The columns in $X$ were arranged in blocks of 10 variables each, each column was normally distributed with zero mean, sample variance $x_j^T x_j / n = 1$ and within-blocks correlation $x_j^T x_l / n = 0.5$. We used the prior settings $\rho = P(\gamma_j = 1) = 1/p$, $a_\phi = l_\phi = 0.01$. The first block had non-zero coefficients $\beta_8 = 0.5, \beta_9 = 0.75, \beta_{10} =$ 1, the second $\beta_{19} = 0.75$, $\beta_{20} = -1$ and $\beta_j = 0$ in all remaining blocks, and $\phi = 1$.

Figure 1 shows the sequence of 50 first models identified by the "Coolblock" algorithm, along with the corresponding $E(\beta \mid y, \gamma, \omega)$ for $p = 100$ (left) and $p = 500$ (right). In both cases the truly active variables were the first to be included: variables 10 and 9 were followed by a joint inclusion of variables 19 and 20 and finally variable 8. The grey lines show model sizes that were skipped by Coolblock. We remark that the bound (14) discarded all these skipped sizes as potential candidates for the posterior mode, except for $|\gamma| = 3$. The model enumeration outlined after Algorithm 2 gave variables 9,10,20 as the best size 3 model, which had posterior probability $4.47 \times 10^{-10}$. In contrast, the posterior mode included variables $9, 10, 19, 20$ and had a posterior probability of $0.505$. Supplementary Figure 3 shows the posterior probabilities for the best model of each size for $p = 100$ and $p = 500$. In the latter $p = 500$ the posterior probability to the data-generating truth (variables 8,9,10,19,20) was roughly 0.9. The Bayesian model averaged means for the regression coefficients (Supplementary Figure 4) exhibit a very strong shrinkage towards 0 for the truly spurious coefficients.

The total runtimes were 2.5 seconds for $p = 100$ and 12.5 seconds for $p = 500$ using the same R language implementation and computer as in the orthogonal example, again we remark that either focusing on the few best models or variables or using more efficient programming would lead to even faster calculations.

### 5.4 Subgroup analysis

Block-diagonal designs occur naturally when we aim to find non-zero regression coefficients within pre-defined groups. For instance, medical treatments may be effective only for certain patient subtypes, or a set of biomarkers may have different effects in different groups. As illustration we take a colon cancer gene expression dataset analysed in [14], selecting 10 genes (ESM1, GAS1, HIC1, CILP, ARL4C, AOC3, URB2, FAM89C, PCGF2, CCDC102B) found potentially related to the response gene TGFB, a growth factor playing an important role in cancer progression. We enumerated all $2^{10}$ models under the prior settings in Section 5.3 and found that the posterior mode included 6 genes: ARL4C, AOC3, URB2, FAM89B, PCGF2 and CCDC102B.

Out of the $n = 262$ patients there are 46, 108 and 108 in cancer stages 1, 2 and 3 respectively (Supplementary file tgfb_data.txt). Since staging is important for cancer progression we explore the dependence of TGFB on the 10 genes within each stage. Formally, this leads to a block-diagonal $X^T X$ with three blocks of 11 variables each (10 genes plus the intercept). Enumerating the $2^{33}$ models is challenging, however to find the posterior mode we only require $\mathcal{O}(2^{11})$ calculations. In stage 1 GAS1 and CILP were selected as having non-zero coefficients, CCDC102B in stage2, and ARL4C, AOC3, URB2 and PCGF2 both in stages 2 and 3. That is, 5 out of the 6 genes were found again in stages 2-3 but this was not the case for FAM89C, further GAS1 (a cancer-related growth supressor) and CILP (associated to local advancement and reactive stroma in breast and prostate cancer) were only found in stage 1, suggesting that TGFB regulation might behave differently in these less aggressive tumours.

### 5.5 Block search for general $X^T X$

An an illustration of the methodology outlined in Section 3 we simulated data with $n = 100$, $p = 500$ and $\beta$ as in Section 5.3. The rows in $X$ were drawn independently from a multivariate Gaussian with mean 0 and covariance $\Sigma$. We set $\sigma_{jj} = 1$ for $j = 1, \ldots, p$ and simulated 1,000 independent datasets under each of the three following covariance structures: compound symmetric where $\sigma_{ij} = 0.5$ for $i \neq j$, autoregressive where $\sigma_{ij} = 0.9^{|i-j|}$ and block-diagonal where $\sigma_{ij} = 0.9$ if variables $i$ and $j$ are in the same block and blocks are defined as in Section 5.3. Under block-diagonal $\Sigma$ the realized $X^T X$ is close to (but not exactly) block-diagonal, whereas compound symmetry leads to a dense $X^T X$ that deviates significantly from block-diagonality. We used Zellner's prior on $\beta_\gamma$ and set $p(\gamma)$ to the Beta-Binomial(1,1) prior.

|  | Block-diagonal | | Autoregressive | | Compound symmetric | |
| --- | --- | --- | --- | --- | --- | --- |
|  | Proportion | Mean | Proportion | Mean | Proportion | Mean |
| $p(\hat{\gamma}_B^* \mid y) = p(\hat{\gamma}_G^* \mid y)$ | 0.79 | -0.29 | 0.85 | -0.17 | 0.80 | -0.56 |
| $p(\hat{\gamma}_B^* \mid y) \geq p(\gamma_T \mid y)$ | 1.00 | 12.43 | 1.00 | 13.66 | 0.89 | 1.26 |
| $p(\hat{\gamma}_G^* \mid y) \geq p(\gamma_T \mid y)$ | 1.00 | 12.72 | 1.00 | 13.83 | 1.00 | 1.82 |

Table 1: Proportion of simulations ($n = 100$, $p = 500$) where each mode has equal or higher posterior probability than another, and mean difference. $\hat{\gamma}_B^*$ and $\hat{\gamma}_G^*$: modes from Block search and Gibbs algorithms. $\gamma_T$: simulation truth.

Table 1 compares the mode $\hat{\gamma}_B^*$ returned by our block search algorithm with $\hat{\gamma}_G^*$, the mode from the computationally-intensive Gibbs sampling algorithm in [11] (function modelSelection in R package mombf). The proportion of simulations in which $p(\hat{\gamma}_B^* \mid y) = p(\hat{\gamma}_G^* \mid y)$ ranged in 0.79-0.85 and the mean difference $\log p(\hat{\gamma}_B^* \mid y) - \log p(\hat{\gamma}_G^* \mid y)$ in [-0.17,-0.56], that is the Bayes factor between $\hat{\gamma}_B^*$ and $\hat{\gamma}_G^*$ gave similar support to both solutions. We also evaluated $p(\gamma_T \mid y)$ where $\gamma_T$ denotes the data-generating truth, finding that $p(\hat{\gamma}_B^* \mid y) \geq p(\gamma_T \mid y)$ in all block-diagonal and autoregressive simulations and 0.88 of the more challenging compound symmetric simulations. The mean $\log p(\hat{\gamma}_B^* \mid y) - \log p(\gamma_T \mid y)$ was positive, indicating that $\hat{\gamma}_B^*$ had higher posterior probability than the simulation truth and hence our block search algorithm did not get stuck in a low-quality local mode. Altogether $\hat{\gamma}_B^*$ provided a computationally-convenient solution that was often optimal or near-optimal.

# 6 Discussion

Whereas there is ample literature on the advantages of selecting variables via best subset selection (see [4] for a recent overview), or analogously in a Bayesian framework by setting positive prior probabilities to zero coefficients, its lack of computational scalability to the size of $p$ has fuelled rapid developments in relaxations of this problem, typically by changing the norm used to penalize complexity (e.g. from $l_0$ to $l_1$ or folded concave). Here we considered a simplified scenario based on assumptions on the design matrix which, in spite of being a classical problem and arising in a number of applications, to our knowledge did not have a fully Bayesian, exact and scalable solution yet. The approach is based on a fast model search algorithm, which incidentally also solves the best subset problem in a frequentist setting when the design matrix is as assumed in this article, and a fast deterministic integration based on expressing an expensive sum over the model space as a convenient product. We also illustrated how these ideas can be exploited for fast model search under general $X^T X$.

Our implementation as currently described can be speeded up. We have used bounds to reduce the effective model space after the application of Coolblock. The idea of using bounds for speeding up best subset selection is old, for example as in the leaps and bounds algorithm of [10], and it has been explored within a broad and powerful optimization toolkit recently, see for example [4]. We are exploring the incorporation of such bounds to trim the model space within our block-search algorithms.

Another direction for generalisation is our assumption that the shrinkage parameter $\tau$ and the variable inclusion parameter $\rho$ are fixed. Although our paradigm can be extended to dealing in a fully Bayesian way with those too, we do not view fixing $\rho$ as problematic, as illustrated in our examples this already offers great control over the degree of prior sparsity and anyway the Bayesian best subset solution for any given $|\gamma|$ does not depend on $\rho$. Regarding the fixed $\tau$ assumption, this is not critical for model search as our algorithms to locate the posterior mode apply directly to any scale mixture over $\tau$, however numerical integration would now be with respect to $\phi$ and $\tau$, increasing somewhat the computational cost.

Finally, our novel approach for identifying the best models of different sizes in the block-diagonal case is only worked out in detail for Zellner's prior. The main convenience is that the important quantities in model selection depend in a simple way on residual sums of squares and model size. These conveniences are exploited by the "Coolblock" algorithm we propose, however one may tackle other priors by finding the roots in terms of $\phi$ of expressions analogous to (13) and using approximations connecting $p(y \mid \gamma, \phi, \omega)$ to $p(y \mid \gamma, \omega)$.

## Acknowledgement

Papaspiliopoulos would like to acknowledge financial support from the Spanish Ministry of Economics via a research grant. Rossell was partially supported National Institute of Health grant R01 CA158113-05 and a Ramón y Cajal fellowship RyC-2015-18544. Both authors thank Mr Miquel Torrens for helpful input in this work.

# Supplement to: Bayesian block-diagonal variable selection and model averaging

O. Papaspiliopoulos

ICREA and Department of Economics and Business, Universitat Pompeu Fabra

D. Rossell

Department of Economics and Business, Universitat Pompeu Fabra *

January 2, 2017

## Abstract

A supplement that includes some experiments not included in the main article and R code to reproduce the experiments. The code requires including the R library mombf.

## Numerical experiments

**Orthogonal design**

We first consider an orthogonal example with $p = 500$, $n = 510$, where as an illustration we set $\rho = P(\gamma_j = 1) = 1/p$, $a_\phi = l_\phi = 0.01$. As a side note the choice $\rho = 1/p$ induces a strong sparsity penalty, in particular stronger than the popular Beta-Binomial(1,1) prior, and leads to the type of exponential penalty found to lead to optimal minimax posterior concentration rates in [1]. The columns in $X$ are normally distributed with zero mean and unit variance and $\beta = (0, \ldots, 0, 0.5, 0.75, 1)$, $\phi = 1$. We consider both Zellner's prior with $\tau = n$ (unit information prior) and the product MOM with the default $\tau = 0.348$ recommended in [2].

Figure 1 (left) shows $p(\gamma \mid y)$ for the sequence of models (up to $|\gamma| = 50$) found by ordering the variables decreasingly in $|x_j^T y|/x_j^T x_j$. The right panel shows $p(\phi \mid y, \omega)$ evaluated on our adaptive grid, which as expected captures the high $p(\phi \mid y, \omega)$ support and is denser in regions where $p(\phi \mid y, \omega)$ changes rapidly.

The posterior mode consists of variables $498, 499, 500$ (the simulation truth) under Zellner's and product MOM priors. For the former the probability at the posterior mode was obtained by appealing to

$$\frac{p(y \mid \gamma, \omega)}{p(y \mid \emptyset, \omega)} = \left( \frac{l_\phi + y^T y}{l_\phi + y^T y - \frac{\tau}{\tau+1} u(y, \gamma)} \right)^{\frac{a_\phi + n}{2}} \frac{1}{(1+\tau)^{|\gamma|/2}}$$

---

*equal contribution by both authors

and for the latter to

$$\frac{p(y \mid \gamma, \omega)}{p(y \mid \emptyset, \omega)} = \left(\frac{l_\phi + y^T y}{l_\phi + y^T y - \frac{\tau n}{\tau n+1} u(y,\gamma)}\right)^{\frac{a_\phi+n}{2}} \frac{1}{(1+\tau n)^{|\gamma|/2}} \times$$

$$\sum_{a_1=0}^{1} \cdots \sum_{a_{|\gamma|}=0}^{1} \frac{\Gamma\left(\frac{a_\phi+n}{2} + |\gamma| - |a|\right)}{\left\{l_\phi + y^T y - \frac{\tau n}{\tau n+1} u(y,\gamma)\right\}^{|\gamma|-|a|} \Gamma\left(\frac{a_\phi+n}{2}\right)}.$$

thus it did not require any additional numerical integration other than $p(y \mid \omega) = \int p(y \mid \phi, \omega) p(\phi \mid \omega)$, which we do at the outset as described in Section 3.5 in the article. Under Zellner's prior both Simpson's rule on our adaptive grid and R function integrate gave $p(y \mid \omega) = e^{-257.7}$, and a resultant posterior probability at the mode of 0.893. For the product MOM we obtained the much higher posterior probability 0.995. Using the already computed $p(y \mid \omega)$ we evaluated $p(\gamma \mid y, \omega)$ for all other $p$ models in the sequence, finding they all had probability less than 0.01 both under Zellner's and the product MOM prior, which by Proposition 1 gives an upper bound on the probability of any other conceivable model. Proposition 1 also gives the next best models of any size. Following the ordering of variables the second best model of size say $l$, is $\gamma_j = 1$ for $j = 1, \ldots, l-1, l+1$ and $\gamma_j = 0$ otherwise, the third best model is either $\gamma_j = 1$ for $j = \ldots, l-1, l+2$ or $j = 1, \ldots, l-2, l, l+1$, and so on. As an example under Zellner's prior the second best model for sizes $2, 3, 4$ combined truly active variables $498, 499, 500$ with the spurious $j = 485$ and all had posterior probability below 0.005.

The grid-based integration in Section 3.5 gave marginal inclusion probabilities and $E(\beta \mid y)$. For the truly active variables $p(\gamma_j = 1 \mid y) = 1$ (up to rounding) under Zellner's and product MOM priors. Figure 2 (left) shows $p(\gamma_j = 1 \mid y)$ for truly spurious variables, whereas the right panel shows $E(\beta_j \mid y)$ versus least-squares estimates. For the active variables $E(\beta_j \mid y) = 0.433, 0.749, 1.065$ under Zellner's and $0.440, 0.751, 1.065$ under the product MOM prior, in both cases fairly close to the data-generating $0.5, 0.75, 1$.

### Block-orthogonal design

Figure 3 shows results for the block-diagonal simulation, analogous to those for the orthogonal design.

## R code

The R code for the orthogonal example is given below.

```
#Simulate data
> library(mombf)
> set.seed(1)
> p <- 500; n <- 510
> x <- scale(matrix(rnorm(n*p),nrow=n,ncol=p),center=TRUE,scale=TRUE)
> S <- cov(x)
> e <- eigen(cov(x))
> x <- t(t(x %*% e$vectors)/sqrt(e$values))
> th <- c(rep(0,p-3),c(.5,.75,1)); phi <- 1
```

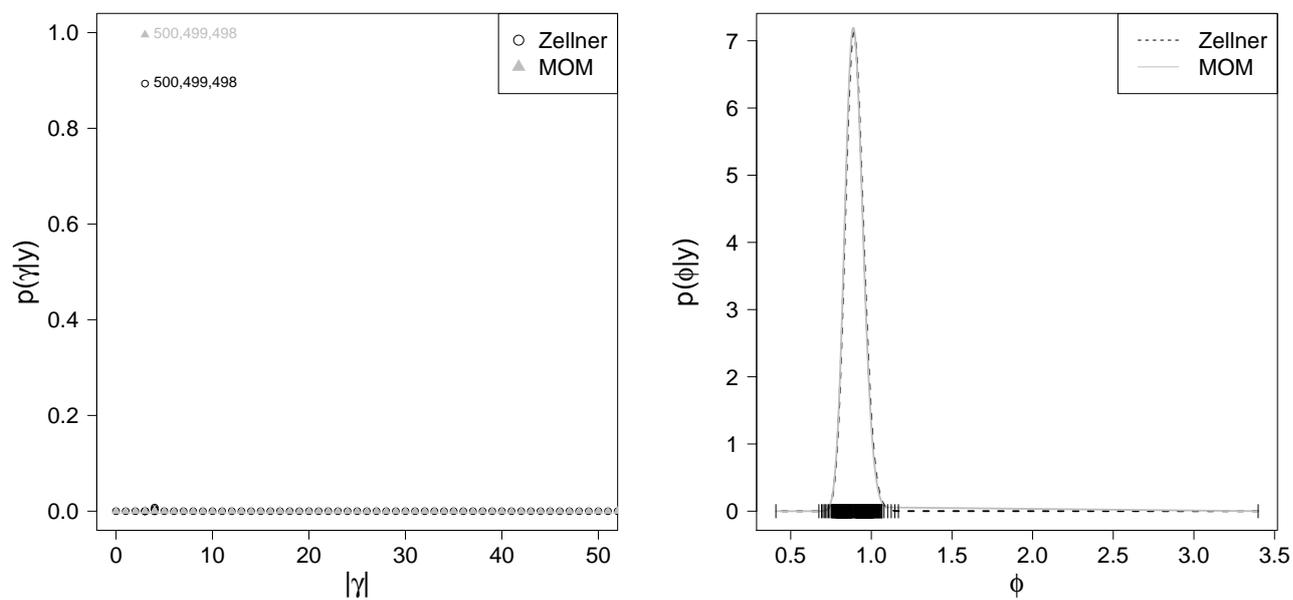

Figure 1: Orthogonal $X^T X$ under Zellner's ($\tau = n = 510$) and MOM ($\tau = 0.348$) priors, $p = 500$. $p(\gamma \mid y, \omega)$ for best models of size $|\gamma| = 0, 1, \ldots, 50$ (left) and $p(\phi \mid y, \omega)$ (right) evaluated on a grid (vertical black segments)

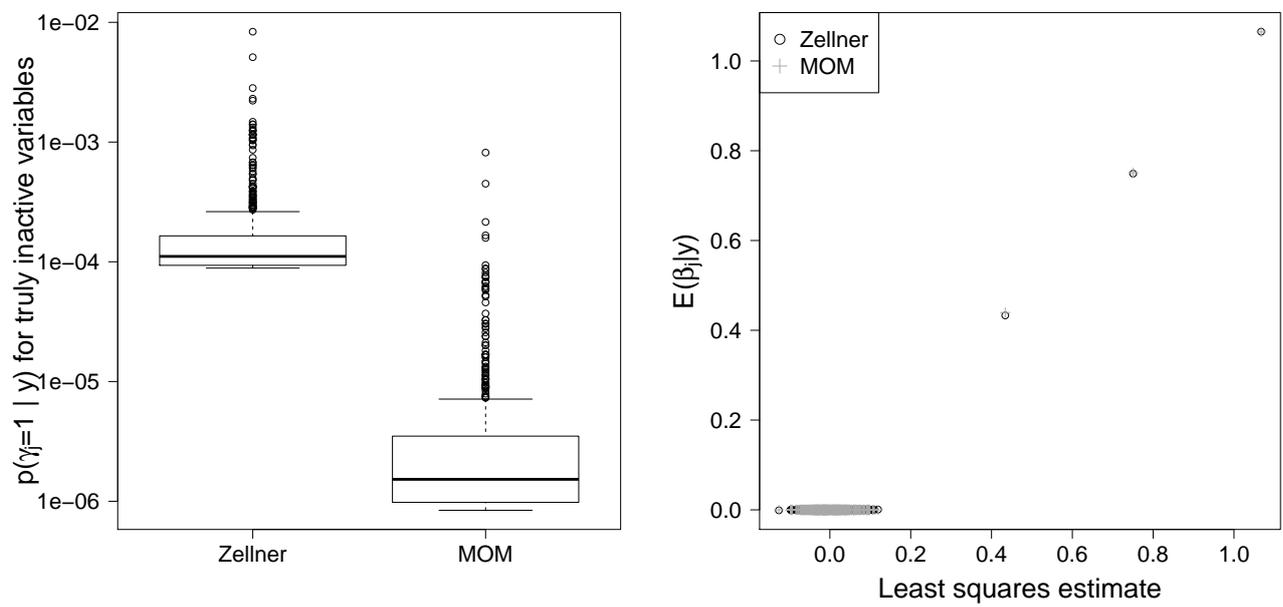

Figure 2: $P(\gamma_j = 1 \mid y, \omega)$ for truly inactive variables (left) and $E(\beta_j \mid y, \omega)$ versus least squares estimate (right) in orthogonal example.

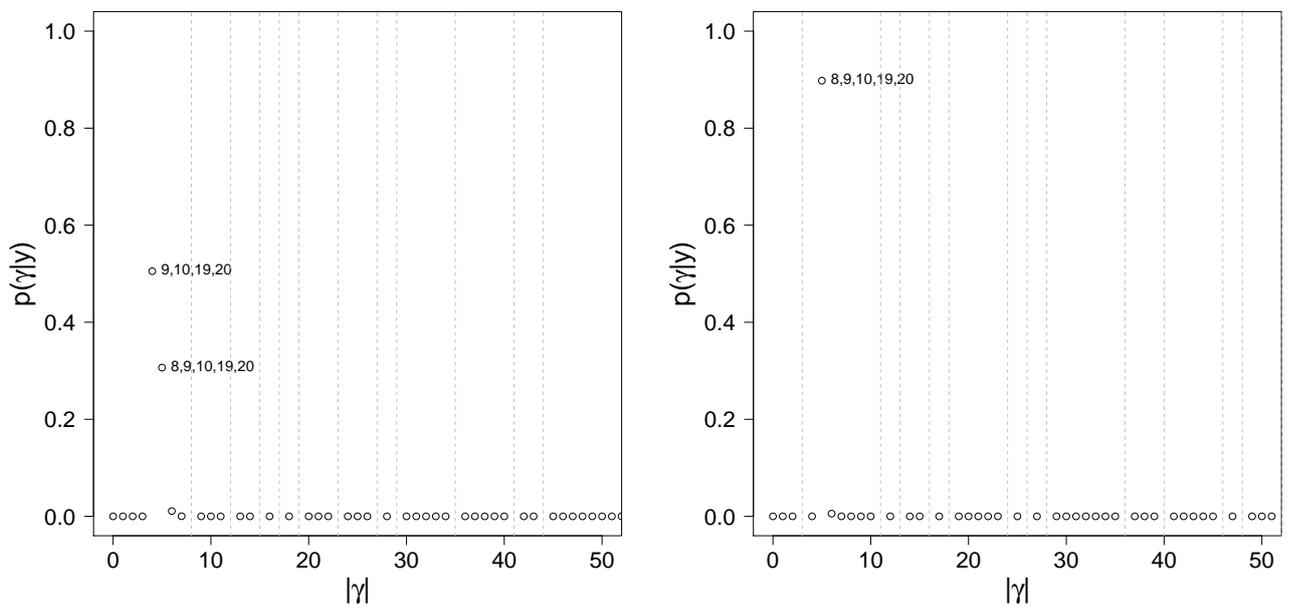

Figure 3: Block-diagonal example: $p(\gamma \mid y, \omega)$ for $p = 100$, $n = 150$ and $p = 500$, $n = 510$

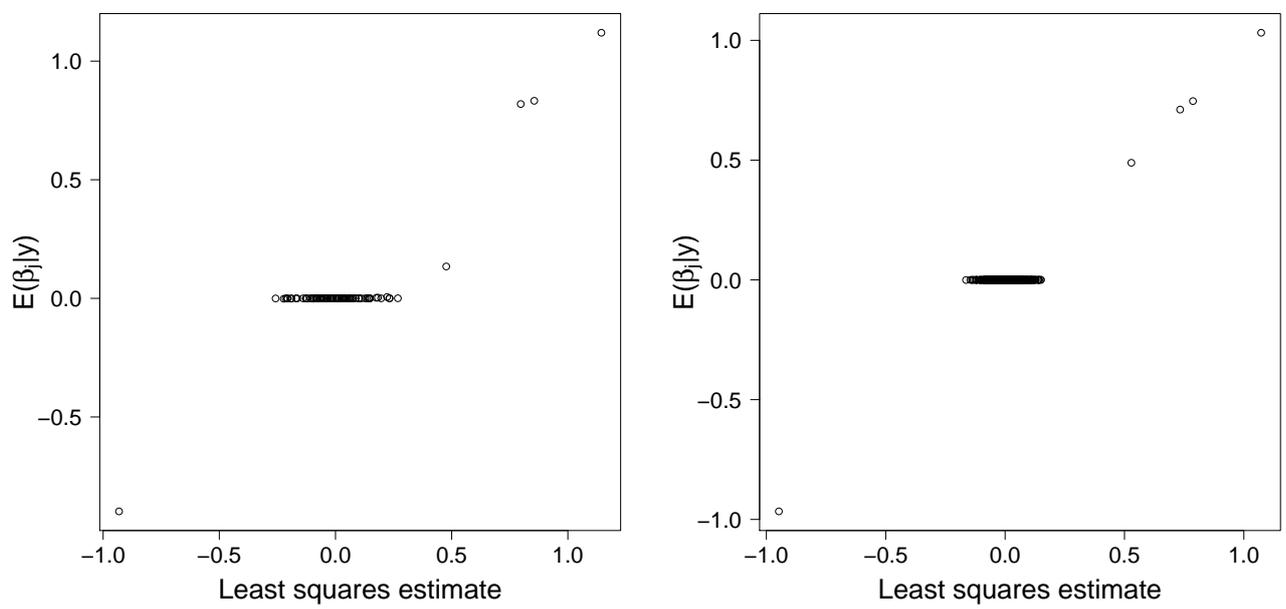

Figure 4: Block-diagonal example: $E(\beta \mid y, \omega)$ for for $p = 100$, $n = 150$ and $p = 500$, $n = 510$

```
> y <- x %*% matrix(th,ncol=1) + rnorm(n,sd=sqrt(phi))

#Fit model
> priorCoef=zellnerprior(tau=n)
> priorDelta=modelbinomprior(p=1/p)
> priorVar=igprior(0.01,0.01)
> pm.zell <-
> postModeOrtho(y,x=x,priorCoef=priorCoef,priorDelta=priorDelta,
priorVar=priorVar,bma=TRUE)
> priorCoef=momprior(tau=0.348)
> pm.mom <- postModeOrtho(y,x=x,priorCoef=priorCoef,priorDelta=priorDelta,
priorVar=priorVar,bma=TRUE)

#Plot posterior model probabilities
> par(mar=c(5,5,1,1))
> nvars <- sapply(strsplit(as.character(pm.zell$models$modelid),split=','),length)
> plot(nvars,pm.zell$models$pp,ylab=expression(paste("p(",gamma,"|y)")),
xlab=expression(paste("|",gamma,"|")),cex.lab=1.5,ylim=0:1,xlim=c(0,50))
> sel <- pm.zell$models$pp>.05
> text(nvars[sel],pm.zell$models$pp[sel],pm.zell$models$modelid[sel],pos=4)
#
> nvars <- sapply(strsplit(as.character(pm.mom$models$modelid),split=','),length)
> points(nvars,pm.mom$models$pp,col='gray',pch=17)
> sel <- pm.mom$models$pp>.05
> text(nvars[sel],pm.mom$models$pp[sel],pm.mom$models$modelid[sel],pos=4,col='gray')
> legend('topright',c('Zellner','MOM'),pch=c(1,17),col=c('black','gray'),cex=1.5)

#Plot posterior of phi
> par(mar=c(5,5,1,1))
>
plot(pm.zell$phi,type='l',xlab=expression(phi),ylab=expression(paste("p(",phi,"|y)")),
cex.lab=1.5,cex.axis=1.2,lty=2,lwd=2)
> points(pm.mom$phi,type='l',col='gray',lwd=2)
> segments(x0=pm.zell$phi[,1],y0=-.1,y1=.1,col=1)
> legend('topright',c('Zellner','MOM'),lty=c(2,1),col=c('black','gray'),cex=1.5)

#Plot BMA estimates
> par(mar=c(5,5,1,1))
> ols <- (t(x) %*% y) / colSums(x^2)
> plot(ols,pm.zell$bma$coef,xlab='Least squares estimate',
ylab=expression(paste('E(',beta[j],'|y)')),cex.lab=1.5,cex.axis=1.2,col=1)
> points(ols,pm.mom$bma$coef,pch=3,col='darkgray')
> legend('topleft',c('Zellner','MOM'),pch=c(1,3),col=c('black','darkgray'))
```

The R code below was used for the block-diagonal example.

```
#Simulate data
> set.seed(1)
> p <- 500; n <- 510
> blocksize <- 10
> blocks <- rep(1:(p/blocksize),each=blocksize)
> x <- scale(matrix(rnorm(n*p),nrow=n,ncol=p),center=TRUE,scale=TRUE)
> S <- cov(x)
> e <- eigen(cov(x))
> x <- t(t(x %*% e$vectors)/sqrt(e$values))
> Sblock <- diag(blocksize)
> Sblock[upper.tri(Sblock)] <- Sblock[lower.tri(Sblock)] <- 0.5
> vv <- eigen(Sblock)$vectors
```

```
> sqSblock <- vv %*% diag(sqrt(eigen(Sblock)$values)) %*% t(vv)
> for (i in 1:(p/blocksize)) x[,blocks==i] <- x[,blocks==i] %*% sqSblock
> th <- rep(0,ncol(x))
> th[blocks==1] <- c(rep(0,blocksize-3),c(.5,.75,1))
> th[blocks==2] <- c(rep(0,blocksize-2),c(.75,-1))
> phi <- 1
> y <- x %*% matrix(th,ncol=1) + rnorm(n,sd=sqrt(phi))

#Fit model
> priorCoef=zellnerprior(tau=n)
> priorDelta=modelbinomprior(p=1/p)
> priorVar=igprior(0.01,0.01)
> pm <- postModeBlockDiag(y=y,x=x,blocks=blocks,priorCoef=priorCoef,
priorDelta=priorDelta,priorVar=priorVar,bma=TRUE)

#Coolblock figure
> maxvars=50
> ylim=range(pm$postmean.model[,-1])
> plot(NA,NA,xlab=expression(paste("|",gamma,"|")),
  ylab=expression(paste('E(',beta[j],'|y,',gamma,')')),
  xlim=c(0,maxvars),ylim=ylim,cex.lab=1.5)
> visited <- which(!is.na(pm$models$pp))
> notvisited <- which(is.na(pm$models$pp))
> for (i in 2:ncol(pm$postmean.model)) {
    lines(pm$models$nvars[visited],pm$postmean.model[visited,i])
  }
> text(maxvars, pm$postmean.model[maxvars,which(th!=0)+1],
  paste('X',which(th!=0),sep=''), pos=3)
> abline(v=pm$models$nvar[notvisited], lty=2, col='gray')
```

The R code for the subgroup analysis is given below.

```
> library(mombf)
> tgfb <- read.table('tgfbdata.txt',header=TRUE,sep='\t')
> y <- tgfb[,1]; stage <- tgfb[,2]; x <- as.matrix(tgfb[,-1:-2])
> p <- ncol(x)+1
> blocks= rep(1:3,each=p)
> xnew= matrix(0,nrow=length(y),ncol=3*p)
> xnew[stage=='1',blocks==1]= cbind(1,x[stage==1,])
> xnew[stage=='2',blocks==2]= cbind(1,x[stage=='2',])
> xnew[stage=='3',blocks==3]= cbind(1,x[stage=='3',])
> colnames(xnew) <- c('int.s1',paste(colnames(x),'s1',sep='.'),'int.s2',
paste(colnames(x),'s2',sep='.'),'int.s3',paste(colnames(x),'s3',sep='.'))

#All patients analysis
> priorCoef=zellnerprior(tau=nrow(x)); priorDelta= modelbinomprior(1/ncol(x))
> ms= modelSelection(y=y,x=x,priorCoef=priorCoef,priorDelta=priorDelta)
> head(postProb(ms))
> bestmodel.noint <- postProb(ms)[1,]
> colnames(x)[as.numeric(strsplit(as.character(bestmodel.noint$modelid),split=',')[[1]])]

#Subgroup analysis
> pm1= postModeBlockDiag(y=y,x=xnew,blocks=blocks,priorCoef=priorCoef,priorDelta=priorDelta,bma=FALSE)
> bestmodel <- pm1$models[which.max(pm1$models$pp),]
> colnames(xnew)[as.numeric(strsplit(as.character(bestmodel$modelid),split=',')[[1]])]
```

For the simulation study to assess the performance of the block search algorithm when $X^T X$ is not truly block-diagonal we used the R code below. We first define a function simModelSearch that runs a single simulation, then call the function repeatedly and summarize the results.

```
simModelSearch <- function(seed,th,phi=1,n,covariance='CS',rho,trueblocksize,blocksize=10,
 priorCoef,priorDelta,priorVar=igprior(.01,.01)) {
  #Simulate data from linear model y= th' x + e, where e ~ N(0,phi) and x ~ N(0,V)
  #V can have compound symmetric, autoregressive or block-diagonal structure
  # Input:
  # - seed: seed for random number generator
  # - th: true regression coefficients
  # - phi: residual variance
  # - n: sample size
  # - covariance: if CS then V[i,j]=rho for all i,j. If AR then V[i,j]=rho^(abs(i-j)).
  #   If 'blockdiag' then V[i,j]=rho within blocks of size trueblocksize, V[i,j]=0 otherwise. Always V[i,i]=1
  # - rho: correlation parameter, see covariance
  # - trueblocksize: if covariance=='blockdiag' this is the block size for the data-generating V.
  #   Ignored if covariance != 'blockdiag'
  # - blocksize: maximum block size in the block screening algorithm
  # - priorCoef: prior on th
  # - priorDelta: prior on the model space
  # - priorVar: prior on phi
  # Output
  # - pmblock: posterior mode found by the block search algorithm and its posterior probability (up
  #   to normalization)
  # - pmmcmc: posterior mode found by the MCMC algorithm in modelSelection and its posterior probability
  #   (up to normalization)
  # - ptrue: simulation truth and its posterior probability (up to normalization)
  p <- length(th)
  if (priorDelta@priorDistr=='binomial' & ('p' %in% names(priorDelta@priorPars))) {
      rho <- priorDelta@priorPars['p']
      priorModel <- function(nvar) nvar*log(rho) + (p-nvar)*log(1-rho)
  } else if (priorDelta@priorDistr=='binomial' & !('p' %in% names(priorDelta@priorPars))) {
      alpha=priorDelta@priorPars['alpha.p']; beta=priorDelta@priorPars['beta.p']
      priorModel <- function(nvar) lbeta(nvar + alpha, p - nvar + beta) - lbeta(alpha, beta)
  } else if (priorDelta@priorDistr=='uniform') {
      rho <- 0.5
      priorModel <- function(nvar) rep(-p*log(2),length(nvar))
  } else { stop("Prior on model space not recognized. Use modelbbprior(), modelunifprior() or modelbinomprior()")

  #Simulate data
  if (covariance=='CS') {   #compound symmetry
      sigma= diag(p); sigma[upper.tri(sigma)]= sigma[lower.tri(sigma)]= rho
  } else if (covariance=='AR') {
      sigma= diag(p); for (i in 2:nrow(sigma)) for (j in 1:(i-1)) sigma[i,j]= sigma[j,i]= rho^(i-j)
  } else if (covariance=='blockdiag') {
      blocks <- rep(1:(p/trueblocksize),each=trueblocksize)
      Sblock <- diag(trueblocksize); Sblock[upper.tri(Sblock)] <- Sblock[lower.tri(Sblock)] <- rho
      sigma <- diag(p); for (i in 1:max(blocks)) sigma[blocks==i,blocks==i]= Sblock
  } else stop("covariance type not recognized")
  set.seed(seed)
  x <- rmvnorm(n,sigma=sigma)
  y <- x %*% matrix(th,ncol=1) + rnorm(n,sd=sqrt(phi))

  #Run block-diagonal search
  ms1 <- modelsearchBlockDiag(y=y,x=x,priorCoef=priorCoef,priorDelta=priorDelta,priorVar=priorVar,
blocksize=blocksize,verbose=FALSE)
  sel1 <- as.numeric(strsplit(ms1$modelid[1],split=',')[[1]])
  f1 <- nlpMarginal(sel=sel1,family='normal',priorCoef=priorCoef,priorVar=priorVar,y=y,x=x,logscale=TRUE)
```

```
+ priorModel(length(sel1))

  #Run MCMC search
  ms2 <- modelSelection(y=y,x=x,center=FALSE,scale=FALSE,priorCoef=priorCoef,priorDelta=priorDelta,
priorVar=priorVar,niter=10^4,method='Laplace',verbose=FALSE)
  pp2 <- postProb(ms2)
  sel2 <- as.numeric(strsplit(as.character(pp2$modelid[1]),split=',')[[1]])
  f2 <- nlpMarginal(sel=sel2,family='normal',priorCoef=priorCoef,priorVar=priorVar,y=y,x=x,logscale=TRUE)
+ priorModel(length(sel2))

  #Evaluate log-posterior for the true model
  seltrue <- which(th!=0)
  ftrue <- nlpMarginal(sel=seltrue,family='normal',priorCoef=priorCoef,priorVar=priorVar,y=y,x=x,logscale=TRUE)
 + priorModel(length(seltrue))
  pmblock <- data.frame(modelsearch='blocksearch',nvars=ms1$nvars[1],modelid=ms1$modelid[1],pp=f1)
  pmmcmc <- data.frame(modelsearch='mcmc',nvars=length(sel2),modelid=pp2[1,'modelid'],pp=f2)
  pmtrue <- data.frame(modelsearch='simtruth',nvars=length(seltrue),modelid=paste(seltrue,collapse=','),pp=ftrue)
  ans <- rbind(pmblock,pmmcmc,pmtrue); rownames(ans) <- NULL
  cat('.')
  return(ans)
}

> library(parallel)
> n <- 100; p <- 500
> th <- c(rep(0,p-12),c(.75,-1),rep(0,7),c(.5,.75,1))
> priorCoef= zellnerprior(tau=n); priorDelta=modelbbprior(1,1); priorVar=igprior(0.01,0.01)

> sim.cs <- mclapply(1:1000, function(i) simModelSearch(seed=i,th=th,phi=1,n=n,covariance='CS',rho=0.5,
blocksize=10,priorCoef=priorCoef,priorDelta=priorDelta,priorVar=priorVar), mc.cores=2)

> sim.ar <- mclapply(1:1000, function(i) simModelSearch(seed=i,th=th,phi=1,n=n,covariance='AR',rho=0.9,
blocksize=10,priorCoef=priorCoef,priorDelta=priorDelta,priorVar=priorVar), mc.cores=2)

> sim.bd <- mclapply(1:1000, function(i) simModelSearch(seed=i,th=th,phi=1,n=n,covariance='blockdiag',
rho=0.9,blocksize=10,trueblocksize=10,priorCoef=priorCoef,priorDelta=priorDelta,priorVar=priorVar),
mc.cores=2)
```